\documentclass[nofootinbib,aps,11pt]{revtex4}
\usepackage{graphicx}
\usepackage{cancel}
\usepackage{amssymb}
\usepackage{textcomp}
\usepackage{amsmath}
\usepackage{bm}
\usepackage{times}
\usepackage{epsfig}
\usepackage{color}
\usepackage{axodraw}
\usepackage{graphicx}
\usepackage{slashed}

\def\lsim{\:\raisebox{-0.5ex}{$\stackrel{\textstyle<}{\sim}$}\:}
\def\gsim{\:\raisebox{-0.5ex}{$\stackrel{\textstyle>}{\sim}$}\:}

\begin{document}

\title{\Large Radiative Neutralino Decay in Split Supersymmetry}
\author{Marco Aurelio D\'\i az}
\author{Boris Panes}
\author{Pedro Urrejola}
\affiliation{
{\small Departamento de F\'\i sica, Universidad Cat\'olica de Chile,
Avenida Vicu\~na Mackenna 4860, Santiago, Chile 
}}
\begin{abstract}
Radiative neutralino decay $\chi^0_2\longrightarrow\chi^0_1\gamma$ is studied 
in a Split Supersymmetric scenario, and compared with mSUGRA and MSSM. This 
1-loop process has a transition amplitude which is often quite small, but has the
advantage of providing a very clear and distinct signature: electromagnetic 
radiation plus missing energy. In Split Supersymmetry this radiative decay is in
direct competition with the tree-level three-body decay 
$\chi^0_2\longrightarrow\chi^0_1 f\overline f$, and we obtain large values for
the branching ratio $B(\chi^0_2\longrightarrow\chi^0_1\gamma)$ which can be 
close to unity in the region $M_2 \sim M_1$. Furthermore, the value for the 
radiative neutralino decay branching ratio has a strong dependence on the
split supersymmetric scale $\widetilde{m}$, which is otherwise very difficult 
to infer from experimental observables.
\end{abstract}

\date{\today}
\maketitle

\section{Introduction}


Split Supersymmetry (SS) was introduced in order to avoid some of the most 
notorious inconveniences of the Minimal Supersymmetric Standard Model (MSSM),
namely, the lack of an automatic mechanism to avoid large flavour changing 
neutral currents and CP violation, and fast proton 
decay \cite{ArkaniHamed:2004fb}. The strategy is to consider all scalars,
with the exception of one Standard Model (SM) like Higgs boson, with a very 
large mass of the order of $\widetilde m$, using unification of gauge 
couplings and the lightest supersymmetric particle (LSP) as Dark Matter 
candidate, as the only guiding 
principles \cite{Giudice:2004tc}. In addition, motivated by the Cosmological
Constant fine tuning problem, the electroweak scale fine tuning is accepted 
as a property of nature to be explained later by other principles to be 
discovered.

In this SS scenario, the light supersymmetric Higgs boson will have SM-like
couplings and will be difficult to differentiate the two models in the absence
of other signals \cite{Wang:2006gp}. The Large Hadron Collider will shortly 
start accelerating protons, and the best chance in this case for the larger
detectors ATLAS \cite{Aad:2009wy} and CMS \cite{:2008zzk} to detect 
supersymmetry is in the chargino and neutralino sector \cite{Polesello:2004aq}.
While the lightest neutralino is the LSP, 
which is stable and candidate to dark matter in this R-Parity conserving model,
the heavier neutralinos will decay into it. In SS, $\chi^0_2$ will not have 
the chance to decay via intermediate scalars, and it will do it via 
intermediate $Z$ bosons, 
$\chi^0_2\longrightarrow\chi^0_1 Z^*\longrightarrow\chi^0_1 f\overline f$.
The other important decay mode is generated at one-loop, the radiative decay
of the neutralino $\chi^0_2\longrightarrow\chi^0_1\gamma$, where all virtual
charged particles contribute inside the loop \cite{Haber:1988px}. This  decay
mode is well studied in the MSSM, and despite being generated at one-loop, 
it can lead to large branching ratios \cite{Ambrosanio:1996gz}.

In this article we are interested in the one-loop two-body decay mode
$\chi^0_2\longrightarrow\chi^0_1\gamma$ in Split Supersymmetry, and its
relative size with respect to the tree-level three-body decay mode
$\chi^0_2\longrightarrow\chi^0_1 f\overline f$. We study the region of 
parameter space where the radiative decay is enhanced, showing it to 
coincide with a relatively wide strip around $M_2\sim M_1$. The signal for 
the radiative decay, an energetic photon plus missing energy, is clean and
experimentally attractive, as long as the photon does not become too soft due 
to lack of phase space. We show that in this strip of parameter space, where
the photon is still easily detectable \cite{Aad:2009wy}, a measurement of the 
two main branching 
ratios can give information on the supersymmetric scale $\widetilde m$.
This is not a small feature because it is very difficult to measure the 
split supersymmetric scale in these models.

\section{Split Supersymmetry}

Above the scale $\widetilde{m}$ the supersymmetric lagrangian is governed by the following R-Parity conserving superpotential,
\begin{equation} 
\label{superpotencial}
W_{MSSM}=
- \lambda^u \widehat{H}_{u}\widehat{Q}\widehat{U}
+ \lambda^d \widehat{H}_{d}\widehat{Q}\widehat{D}
+ \lambda^e \widehat{H}_{d}\widehat{L}\widehat{E}
- \mu\widehat{H}_{u}\widehat{H}_{d}
\end{equation}
where $\lambda^u$, $\lambda^d$, and $\lambda^e$ are the Yukawa coupling 
$3\times 3$ matrices, and $\mu$ is the Higgs supersymmetric mass parameter. 
>From this superpotential we highlight the following terms in the MSSM
lagrangian,
\begin{eqnarray}
\label{MSSM_lagrangian}
\mathcal{L}_{MSSM} & = & 
-\displaystyle{\frac{g^2}{8}}(H_u^{\dagger}{\sigma}^a H_u + 
H_d^{\dagger}{\sigma}^a H_d)^2 - \displaystyle{\frac{g'^{2}}{8}}
\left(H_u^{\dagger} H_u - H_d^{\dagger} H_d \right)^2 \ 
\nonumber\\  &&
-(m_{H_u}^2+\mu^2)H_u^{\dagger} H_u - (m_{H_d}^2+\mu^2)H_d^{\dagger} H_d
+m_{H_{ud}}^2(H_u^T\epsilon H_u+h.c.)
\nonumber\\  &&
+\lambda^u_{ij}H_u^T\epsilon\bar{u}_i q_j - 
\lambda^d_{ij} H_d^T \epsilon \bar{d}_i q_j 
- \lambda^e_{ij} H_d^T \epsilon \bar{e}_i \ell_j \ 
\\ &&
- \displaystyle{\frac{H_u^{\dagger}}{\sqrt{2}}} 
\big(g{\sigma}^a\widetilde{W}{}^a + 
g' \widetilde{B} \big) \widetilde{H}_u - 
\displaystyle{\frac{H_d^{\dagger}}{\sqrt{2}}}
\big(g {\sigma}^{a} \widetilde{W}{}^a + 
g'\widetilde{B} \big) \widetilde H_d + \mathrm{h.c.}
\nonumber
\end{eqnarray}
where $\epsilon = i \sigma_2$. This lagrangian is valid at the scale
$\widetilde m$ and above. At $\widetilde m$ the Higgs potential is 
characterized by quadratic terms proportional to squared gauge coupling 
constants, $g(\widetilde m)$ and $g'(\widetilde m)$, plus three mass 
terms. The two Higgs eigenstates are found to be rotations of $H_u$ and 
$H_d$ by an angle $\beta$, the lightest one given by 
$H=-\cos{\beta}\epsilon H_d^{*}+\sin{\beta}H_u$.
Also in the MSSM lagrangian we have the Yukawa interactions with
couplings $\lambda^u_{ij}(\widetilde m)$, $\lambda^d_{ij}(\widetilde m)$, 
and $\lambda^e_{ij}(\widetilde m)$. Finally, we see that the 
higgsino-Higgs-gaugino vertex are proportional to the gauge couplings, as 
Higgs-Higgs-gauge boson couplings are, as dictated by supersymmetry.

The Split supersymmetric lagrangian, valid at the scale $\widetilde m$ and
below, is given by, \cite{Giudice:2004tc}
\begin{eqnarray}
\label{SS_lagrangian}
\mathcal{L}_{SS} & = & m^2HH^{\dagger} - \displaystyle{\frac{\lambda}{2}}
\left( H^{\dagger} H \right){}^2
-\Big[ h^u_{ij} \bar{q}_j u_i \epsilon H^* + h^d_{ij} \bar{q}_j d_i H + 
h^e_{ij} \bar{\ell}_j e_i H +{} \nonumber\\ 
&& {} + \displaystyle{\frac{H^{\dagger}}{\sqrt{2}}} 
\big(\widetilde{g}_u {\sigma}^a 
\widetilde{W}{}^a + \widetilde{g}'_u \widetilde B \big) \widetilde{H}_u
+ \displaystyle{\frac{ H^T \epsilon}{\sqrt{2}}}
\big(-\widetilde{g}_d {\sigma}^{a} \widetilde{W}{}^a + \widetilde{g}'_d
\widetilde{B} \big) \widetilde H_d + \mathrm{h.c.} \Big]
\end{eqnarray}
where the Higgs field $H$ is the surviving Higgs doublet at low energies. 
The Higgs potential is defined by a mass term $m^2$ and a quartic self 
coupling $\lambda$. The electroweak symmetry breaking occurs since
$m^2>0$, and the Higgs field acquires a vacuum expectation value $v$. 
The matching condition for the Higgs self interaction at the split
supersymmetric scale $\widetilde m$ is
\begin{equation}
\lambda(\widetilde{m}) = \displaystyle{\frac{g^2(\widetilde{m})+
g'^2(\widetilde{m})}{4}}\cos^2{2\beta}, 
\label{LambdaMatch}
\end{equation}
and this coupling should be run down to the weak scale to find the correct 
electroweak
symmetry breaking. The Yukawa couplings in the split sumersymmetric model
are called $h^u_{ij}$, $h^d_{ij}$, and $h^e_{ij}$, and at the scale 
$\widetilde m$ the corresponding matching condition are
\begin{equation}
h^u_{ij}(\widetilde{m}) = \lambda^{u}_{ij}(\widetilde{m})\sin{\beta},
\qquad
h^{e,d}_{ij}(\widetilde{m}) = \lambda^{e,d}_{ij}(\widetilde{m})\cos{\beta}.
\end{equation}
Finally, we notice from the Split Supersymmetric lagrangian in 
eq.~(\ref{SS_lagrangian}) the Higgs-gaugino-higgsino interactions, whose 
couplings have the following matching conditions with the analogous 
terms in the MSSM lagrangian of eq.~(\ref{MSSM_lagrangian}),
\begin{eqnarray}
\widetilde{g}_u(\widetilde{m}) = g(\widetilde{m})\sin{\beta},
&\qquad &
\widetilde{g}_d(\widetilde{m}) = g(\widetilde{m})\cos{\beta}, 
\nonumber \\
\widetilde{g}'_u(\widetilde{m}) = g'(\widetilde{m})\sin{\beta},
&\qquad &
\widetilde{g}'_d(\widetilde{m}) = g'(\widetilde{m})\cos{\beta}, 
\end{eqnarray}
\begin{figure}[!hhh]
\begin{center}
\includegraphics[height=10cm]{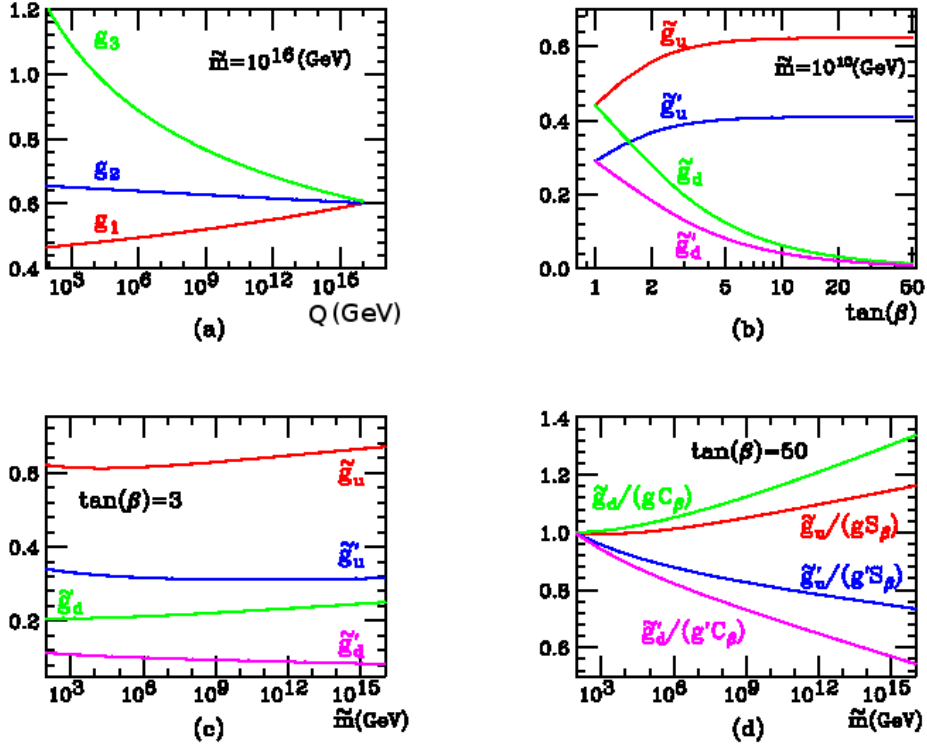}
\vspace{-3ex} \caption{(a) Gauge coupling unification is preserved in 
Split-SUSY. (b)
 Split-Susy couplings dependence on $\tan\beta$. (c) Split-Susy couplings
 evaluated at weak scale as function of $\tilde{m}$. (d) Split-Susy 
 couplings normalized with gauge couplings evaluated at 
 weak scale, for different values of $\tilde{m}$.}
\label{rges}
\end{center}
\end{figure} 
The renormalization group equations for these and other couplings can be 
found in ref.~\cite{Giudice:2004tc}. For illustration we show in Fig.~\ref{rges}
some of their behaviour. In Fig~\ref{rges}(a) we have the running of the 
gauge coupling constants in Split Supersymmetry, which unify at the GUT 
scale as in the MSSM. In the second frame Fig.~\ref{rges}(b) we plot the 
Higgs-gaugino-higgsino couplings as a function of $\tan\beta$ for
$\widetilde m=10^{16}$ GeV. If $\tan\beta=1$, neither the boundary 
condition nor the RGE differentiate between $\widetilde{g}_u$ and 
$\widetilde{g}_d$ or between $\widetilde{g}'_u$ and $\widetilde{g}'_d$.
A sharp splitting appears when $\tan\beta$ increases. The down couplings 
become smaller than $0.1$ for $\tan\beta\gsim 10$, while the up couplings 
approach asymptotically a maximum as $\tan\beta$ increases. In 
Fig.~\ref{rges}(c) we plot the Higgs-gaugino-higgsino couplings as a function
of the split supersymmetric scale $\widetilde m$. In Fig.~\ref{rges}(d) we plot 
the Higgs-gaugino-higgsino couplings normalized by the gauge couplings evaluated 
at the weak scale as a function of the scale $\widetilde{m}$. We choose the value 
$\tan\beta=50$, and observe deviations up to $\pm20\%$. Of course, if
the split supersymmetric scale is taken equal to the weak scale, there 
is no deviation.

Now we introduce the following notation,
\begin{eqnarray}
\tan{\widetilde\beta} \ &\equiv & \ 
\displaystyle{\frac{\widetilde{g}_u}{\widetilde{g}_d}}
\bigg|_{m_W} \hspace{-2ex}
\simeq \tan{\beta}\left[1+
\displaystyle{\frac{\cos{2\beta}}
{64\pi^2}}\Big( 7g^2(\widetilde{m})-3g'^2(\widetilde{m}) 
\Big) \ln\Big(\frac{\widetilde{m}}{m_W}\Big) \right]\\
\tan{\widetilde\beta'} \ &\equiv & \ 
\displaystyle{\frac{\widetilde{g}_u'}{\widetilde{g}_d'}}
\bigg|_{m_W} \hspace{-2ex}
\simeq \tan{\beta}\left[1-
\displaystyle{\frac{\cos{2\beta}}
{64\pi^2}}\Big( 9g^2(\widetilde{m})+3g'^2(\widetilde{m}) 
\Big) \ln\Big(\frac{\widetilde{m}}{m_W}\Big) \right]
\label{appTanBetas}
\end{eqnarray}
where it is understood that $\beta$ is defined at the scale 
$\widetilde{m}$, while $\widetilde\beta$ and $\widetilde\beta'$ are 
defined at the weak scale. The approximated expressions in 
eq.~(\ref{appTanBetas}) is obtained from the corresponding RGE. These 
definitions together with,
\begin{equation}
\widetilde{g}^2 \equiv \widetilde{g}_u^2(m_W) + \widetilde{g}_d^2(m_W), 
\qquad 
\widetilde{g}'{}^2 \equiv \widetilde{g}'_u{}^2(m_W) + 
\widetilde{g}'_d{}^2(m_W),
\end{equation}
allow us to write the neutralino and chargino mass matrices in such a way it
resembles those of the MSSM.
\begin{figure}[!hhh]
\begin{center}
\includegraphics[height=5cm]{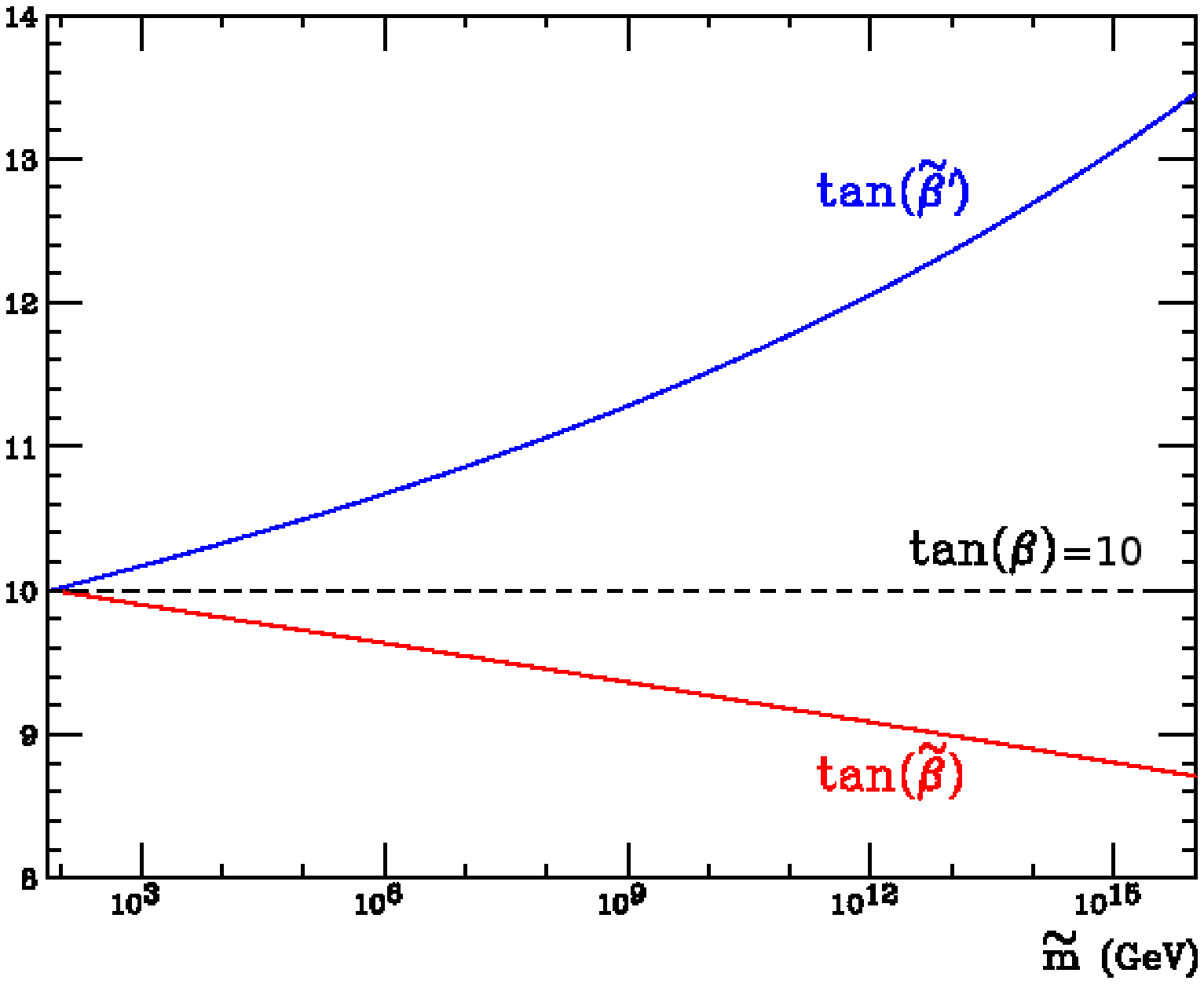}
\includegraphics[height=5cm]{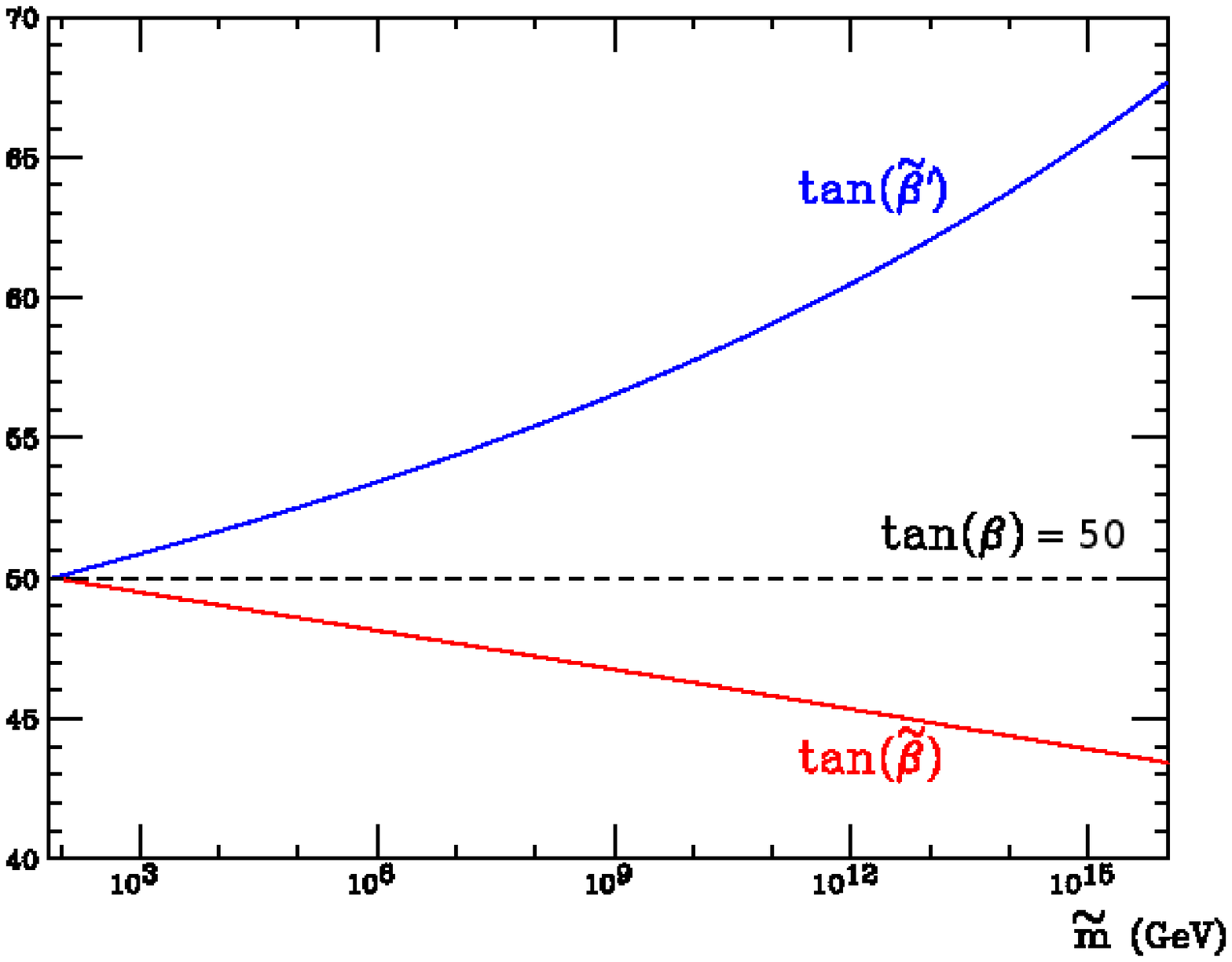}
\vspace{-3ex} \caption{Dependence of $\tan{\widetilde\beta}$ and 
$\tan{\widetilde\beta^{\prime}}$ on the Split Supersymmetric scale, for 
different values of $\tan{\beta}$. }
\label{tbetas}
\end{center}
\end{figure}
The mixing angles $\widetilde\beta$ and $\widetilde\beta'$ are plotted in
Fig.~\ref{tbetas} as a function of $\widetilde m$ for $\tan\beta=10, 50$. Of 
course, there is no difference between the three angles if 
$\widetilde m=M_{weak}$.

With this notation, the neutralino mass matrix in Split Supersymmetric models 
has the following form,
\begin{equation}
{\bf M}_{\chi^0}^{SS}=\left[\begin{array}{cccc}
M_1 & 0 & -\frac{1}{2}\tilde g' v  \tilde c'_\beta & 
\frac{1}{2}\tilde g' v \tilde s'_\beta \\
0 & M_2 & \frac{1}{2}\tilde g v \tilde c_\beta & 
-\frac{1}{2}\tilde g v \tilde s_\beta \\
-\frac{1}{2}\tilde g' v \tilde c'_\beta & 
\frac{1}{2}\tilde g v  \tilde c_\beta & 0 & -\mu \\
\frac{1}{2}\tilde g' v \tilde s'_\beta & 
-\frac{1}{2}\tilde g v \tilde s_\beta & -\mu & 0
\end{array}\right]
\label{X0massmat}
\end{equation}
which is written in the usual basis $\psi^0=(\widetilde{B},
\widetilde{W}^3,\widetilde{H}_{d}^{0},\widetilde{H}_{u}^{0})$. We have used 
the notation $\tilde c'_\beta\equiv\cos\tilde\beta'$, 
$\tilde c_\beta\equiv\cos\tilde\beta$, $\tilde s'_\beta\equiv\sin\tilde\beta'$,
$\tilde s_\beta\equiv\sin\tilde\beta$. This mass matrix is
diagonalized by the matrix $N$, such that 
$N^* {\bf M}_{\chi^0}^{SS} N^{-1} = ({\bf M}_{\chi^0}^{SS})_{diag}$,
and the eigenvectors 
$\widetilde{\chi}_i^0 = N_{ij} {\psi}_j^0$ are the neutralinos.

\section{Neutralino Decays in mSUGRA and Split Supersymmetry}

%
\begin{figure}[!hhh]
\begin{center}
\includegraphics[angle=0,height=4cm]{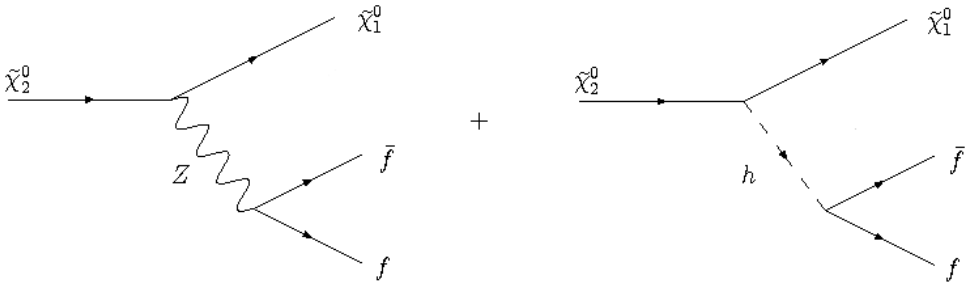}
\vspace{-3ex}\caption{Tree-level three body decay diagrams in Split-SUSY
for the second lightest neutralino.}
\label{three_body_SS}
\end{center}
\end{figure} 
In Split Supersymmetry the three body decay modes of the second lightest
neutralino receive contributions from intermediate gauge and light Higgs 
bosons, with negligible contribution from sfermions and heavy Higgs boson. 
These graphs are in Fig.~\ref{three_body_SS}, where the major contribution 
is from the $Z$-boson exchange, since the fermions in the final 
states have a very small mass. We calculate these decay rates 
integrating over the phase space with numerical techniques. We compare 
our calculations for the case of a small SS scale $\widetilde m\sim 1$ 
TeV with results from the ISASUGRA code \cite{Paige:1998ux} with $m_0\sim 1$ 
TeV. These are in agreement within small differences, the main of which 
is the distinctive running of Higgs-higgsino-gaugino couplings present is SS.

\begin{figure}[!hhh]
\begin{center}
\includegraphics[angle=0,height=3.8cm]{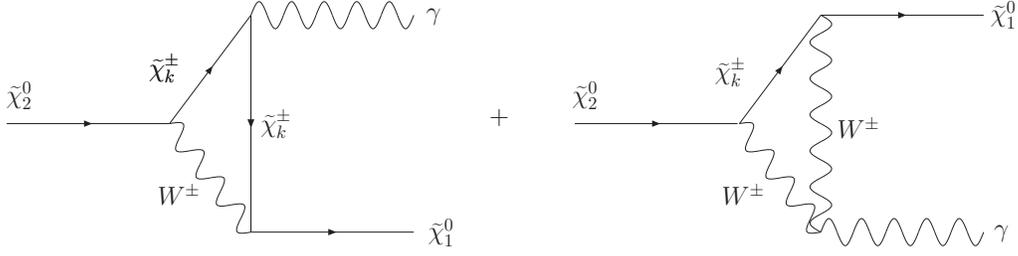}
\vspace{-3ex}\caption{Diagrams for the second lightest neutralino radiative 
decay into a photon and the lightest neutralino in Split-SUSY.}
\label{one_loop_SS}
\end{center}
\end{figure}
Our main interest in this article is the radiative decay of the second 
lightest neutralino into the LSP and a photon. This one-loop generated decay 
can shed light into the properties of heavy particles present in the loop,
charginos in the case of Split Supersymmetry. In addition, it is an 
experimentally interesting decay mode since includes only a hard photon plus 
missing energy. Contributing loops in SS are displayed in Fig.~\ref{one_loop_SS}.
The loops include both charginos and $W$ gauge bosons (charged Goldstone 
bosons are implicit). All other charged scalars which could contribute have
a mass of the order of $\widetilde m$ and they are neglected. Of
course, the effect of the heavy particles is felt via the RGE of the 
effective couplings below $\widetilde m$. We calculate the integral over 
internal momenta analytically using dilogarithms \cite{Haber:1988px}.

\begin{figure}[!hhh]
\begin{center}
\includegraphics[angle=0,width=16cm]{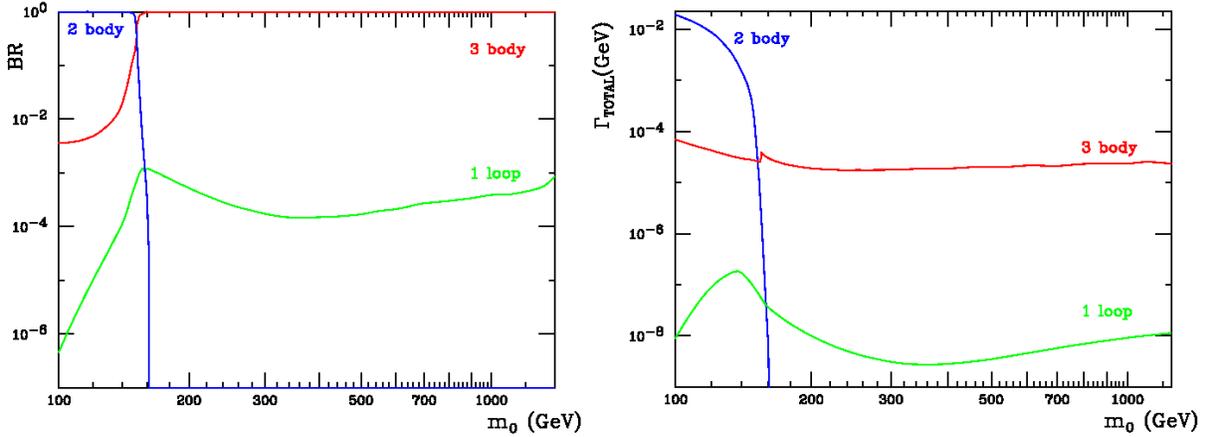}
\vspace{-3ex}\caption{Branching ratio and decay width for the different 
neutralino decay modes, in mSUGRA as a function of $m_0$ and around the 
SPS1a benchmark point.} 
\label{BR_Gamma_m0}
\end{center}
\end{figure}
In Fig.~\ref{BR_Gamma_m0} we show the behaviour of the three main decay 
modes of the second lightest neutralino in mSUGRA: the tree-level two-body 
decay $\chi^0_2\rightarrow \tilde\ell^\pm \ell^\mp$, the tree-level 
three-body decay $\chi^0_2\rightarrow\chi^0_1 f\overline f$, and the one-loop 
two-body decay $\chi^0_2\rightarrow\chi^0_1\gamma$. We have Branching Ratios 
to the left and decay rates to the right, as a function of the universal 
scalar mass $m_0$. The tree-level modes are calculated with the code ISASUGRA,
while the one-loop mode with our code. The other parameters are taken as in 
benchmark SPS1a \cite{snowmass}, which is given in Table \ref{tab:MSSMabc}.
\begin{table}
\begin{center}
\caption{Input parameters for SPS1a mSUGRA benchmark point.}
\bigskip
\begin{minipage}[t]{0.8\textwidth}
\begin{tabular}{ccc}
\hline
Parameter & Value & Units \\
\hline \hline
$m_0$  & 100 & GeV  \\
$M_{1/2}$ & 250 & GeV  \\
$A_0$  & 100 & GeV  \\
$\tan\beta$ & 10 & -  \\
${\mathrm{sign}}(\mu)$ & +1 & -  \\

\hline \hline \label{tab:MSSMabc}
\end{tabular}
\end{minipage}
\end{center}
\end{table}

We can see that the one-loop decay rate 
$\Gamma(\chi^0_2\rightarrow\chi^0_1 \gamma)$ has a maximum near $m_0=150$ 
GeV, decreasing for larger universal scalar mass. At values above $m_0=1$ 
TeV the mSUGRA result differs between 3-20\% from our own SS result,
calculated with values $\widetilde m\gsim 1$ TeV. Of course, it is expected 
that mSUGRA results approach the SS result for large $m_0$, since at large 
values of the universal scalar mass the triangular contributions from heavy 
scalar particles diminish. In SS the effect of these heavy particles appear 
through the RGE of the different couplings, but small differences remain 
between the two approaches because the running couplings include leading 
logarithm effects from all loops. We remind the reader that these RGE 
effects in SS do not spoil the unification of gauge coupling constants, as 
stressed in ref.~\cite{Giudice:2004tc} and illustrated in Fig.~\ref{rges}.
The branching ratio $B(\chi^0_2\rightarrow\chi^0_1 \gamma)$ remains between
$10^{-3}$ and $10^{-4}$ for this mSUGRA scenario.
\begin{table}
\begin{center}
\caption{Chargino and Neutralino masses for SPS1a.}
\bigskip
\begin{minipage}[t]{0.8\textwidth}
\begin{tabular}{ccc}
\hline
Particle & Mass & Units \\
\hline \hline
$\tilde\chi^0_1$  &  99 & GeV  \\
$\tilde\chi^0_2$  & 175 & GeV  \\
$\tilde\chi^0_3$  & 352 & GeV  \\
$\tilde\chi^0_4$  & 372 & GeV  \\
\hline
$\tilde\chi^+_1$  & 175 & GeV  \\
$\tilde\chi^+_2$  & 372 & GeV  \\

\hline \hline \label{tab:chaneu}
\end{tabular}
\end{minipage}
\end{center}
\end{table}
In Table \ref{tab:chaneu} we show the neutralino and chargino masses for
$m_0=100$ GeV, with masses only slightly increasing (one or two GeV) for 
larger scalar mass, calculated using SUSPECT \cite{Djouadi:2002ze}.

We compare in Fig.~\ref{BR_Gamma_m0} the one-loop generated decay mode
$\chi^0_2\rightarrow\chi^0_1 \gamma$, with the tree-level decay modes. 
We define the tree-level two-body decay 
$\chi^0_2\rightarrow \tilde\ell^\pm \ell^\mp$ as the sum of the three 
leptonic decays
$\chi^0_2\rightarrow \tilde e e, \tilde\mu \mu, \tilde\tau \tau$ which 
occurs for values $m_0\lsim 160$ GeV, where the sleptons have a mass 
smaller than $m_{\chi^0_2}$. In this region, this decay mode dominates with 
a branching ratio near unity. We also have the tree-level three-body decay 
$\chi^0_2\rightarrow\chi^0_1 f\overline f$, where we sum over all possible 
fermions. Above $m_0\sim 160$ GeV the off-shell intermediate particles 
which contribute are the $Z$ gauge boson and the squarks or sleptons, 
depending whether the final state fermion is a quark or a lepton. In this 
region the $B(\chi^0_2\rightarrow\chi^0_1 f\overline f)$ is near unity. 
Below $m_0\sim 160$ GeV the contribution from the intermediate light 
on-shell sfermion is removed, and 
$B(\chi^0_2\rightarrow\chi^0_1 f\overline f)$ drops to a value between 
$10^{-3}$ and $10^{-2}$.
\begin{table}
\begin{center}
\caption{Slepton and squark masses for SPS1a.}
\bigskip
\begin{minipage}[t]{0.8\textwidth}
\begin{tabular}{ccc}
\hline
Particle & Mass & Units \\
\hline \hline
$\tilde e_R$, $\tilde\mu_R$  & 145 & GeV  \\
$\tilde e_L$, $\tilde\mu_L$  & 204 & GeV  \\
$\tilde\tau_1$  & 136 & GeV  \\
$\tilde\tau_2$  & 208 & GeV  \\
\hline
$\tilde t_1$  & 375 & GeV  \\
$\tilde b_1$  & 491 & GeV  \\

\hline \hline \label{tab:slepsqu}
\end{tabular}
\end{minipage}
\end{center}
\end{table}
In Table \ref{tab:slepsqu} we show the slepton and the lightest squark 
masses for the SPS1a scenario using SUSPECT. For larger $m_0$ these masses 
grow up sharply, with right selectron and smuon becoming on-shell if 
$m_0\lsim 150$ GeV and similarly for the lightest stau if $m_0\lsim 155$
GeV. Both thresholds are fused into one in Fig.~\ref{BR_Gamma_m0} because
of the low resolution used in the graph.

\begin{figure}[!hhh]
\begin{center}
\includegraphics[angle=0,width=16cm]{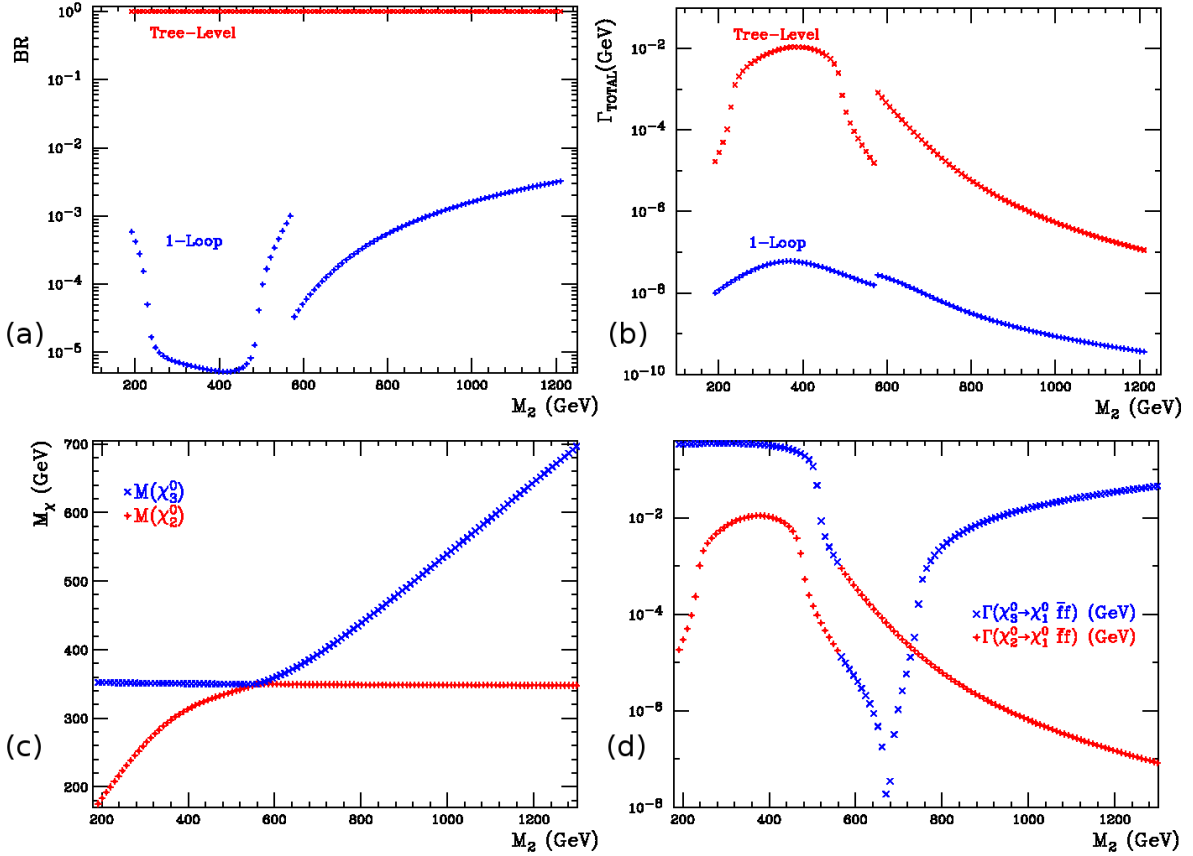}
\vspace{-3ex}\caption{Branching ratio (a) and decay width (b) for 
$\tilde\chi^0_2$ decay modes in SS as a function of $M_2$. In (c) we see 
the mass eigenvalues crossing between the second and third neutralinos, 
which cause the discontinuous behaviour of the branching ratio and decay 
width (d).} 
\label{BR_Gamma_Masas_M2}
\end{center}
\end{figure}
In Fig.~\ref{BR_Gamma_Masas_M2} we show the tree-level three-body decay 
$\chi^0_2\rightarrow\chi^0_1 f\overline f$ and the one-loop two-body
decay $\chi^0_2\rightarrow\chi^0_1 \gamma$ in Split Supersymmetry, as a
function of the wino mass $M_2$, both calculated with our own code.
\begin{table}
\begin{center}
\caption{Input parameters for Split Supersymmetry benchmark point.}
\bigskip
\begin{minipage}[t]{0.8\textwidth}
\begin{tabular}{ccc}
\hline
Parameter & Value & Units \\
\hline \hline
$M_1$  & 102 & GeV  \\
$M_2$  & 192 & GeV  \\
$M_3$  & 587 & GeV  \\
$\mu$  & 357 & GeV  \\
$\tan\beta$ & 10 & -  \\

\hline \hline \label{tab:M1M2mu}
\end{tabular}
\end{minipage}
\end{center}
\end{table}
We choose as SS benchmark point the one given in Table \ref{tab:M1M2mu},
whose soft gaugino and higgsino mass values coincide with the low energy 
soft masses from SPS1a. When varying the wino mass $M_2$ in
Fig.~\ref{BR_Gamma_Masas_M2}, we vary also the bino mass $M_1$ keeping 
constant the $M_2/M_1$ ratio as in our SS benchmark scenario. In frame 
(a) we have the branching ratios, where tree-level three-body decay 
dominates over the one-loop decay $\chi^0_2\rightarrow\chi^0_1 \gamma$ 
with a BR near unity. Since we work in SS, the 
$\chi^0_2\rightarrow\chi^0_1 f\overline f$ mode is mediated only by an 
intermediate $Z$ gauge boson. Similarly, the 
$\chi^0_2\rightarrow\chi^0_1 \gamma$ decay is generated only by quantum 
corrections with $W$ gauge bosons and charginos inside the loop. 

In frame (b) we plot the decay rates for these two modes. The discontinuity 
on both decay rates occur near $M_2\sim 560$ GeV and corresponds to an 
eigenvalue crossing. In frame (c) we see this eigenvalue crossing, with
a $\tilde\chi^0_3$ higgsino type and $\tilde\chi^0_2$ gaugino type for
$M_2\lsim 560$, while the opposite occurs for $M_2\gsim 560$. The effect
of the crossing can be seen very clearly in frame (d) where we have the 
three-body decays for both neutralinos $\tilde\chi^0_2$ and $\tilde\chi^0_3$.

\section{High Radiative Neutralino Decay Branching Ratio} 

In this chapter we analyze with more detail the radiative decay for the 
second lightest neutralino, and look for conditions for an enhanced
$B(\chi^0_2\rightarrow\chi^0_1 \gamma)$ in Split Supersymmetry.
\begin{figure}[!hhh]
\begin{center}
\includegraphics[angle=0,height=10.0cm]{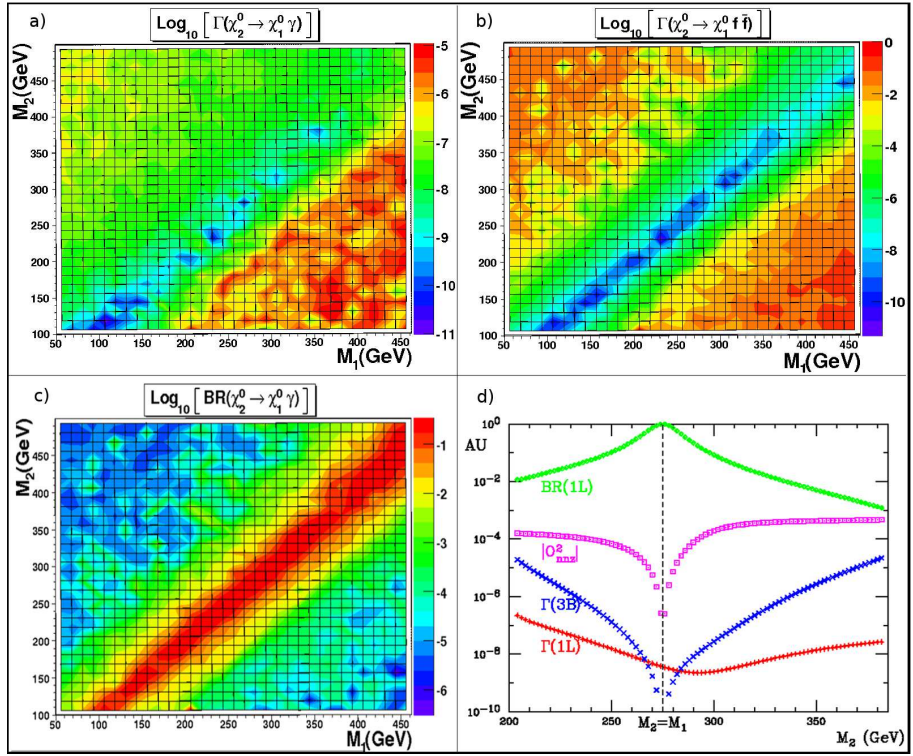}
\vspace{-3ex}\caption{(a) Radiative decay width, (b) 3-Body decay width, 
and (c) Branching ratio for the radiative mode in $M_1$-$M_2$ plane
in Split Supersymmetry. In (d) we have different curves as a function of
$M_2$ showing the special behaviour at $M_1=M_2$.}
\label{M1M2Scan}
\end{center}
\end{figure}
In Fig.~\ref{M1M2Scan}a we show with a color code the logarithmic values for 
the decay rate $\Gamma(\chi^0_2\rightarrow\chi^0_1 \gamma)$ in the gaugino 
mass plane $M_1$-$M_2$, with $\Gamma$ measured in GeV. We show gaugino 
masses smaller than 500 GeV, and vary randomly the values for $\mu$ and 
$\tan\beta$. The largest values for the decay rate occur for $M_1>M_2$, 
reaching typically up to $10^{-5}$ GeV. In the opposite case, when $M_1<M_2$, 
the decay ratio varies typically between $10^{-7}$ and $10^{-8}$ GeV. There 
is a narrow fissure around $M_1\sim M_2$ where the decay rate drops to values 
between $10^{-10}$ and $10^{-11}$ GeV. The fissure is not exactly at the 
bisector but somewhat deviated to the $M_1>M_2$ side of the quadrant. In 
this fissure $\Delta m_\chi\equiv m_{\chi^0_2}-m_{\chi^0_1}$ is minimum,
and the drop of the decay rate is a kinematical suppression.

In Fig.~\ref{M1M2Scan}b we have a similar plot for the decay rate
$\Gamma(\chi^0_2\rightarrow\chi^0_1 f\overline f)$, which is much larger than 
the previous case. This decay rate is more symmetrical with respect to the 
bisector, with decay rates reaching maximum values between $10^{-2}$ and 1 
GeV, and a deep fissure at $M_1\sim M_2$ going all the way down to $10^{-11}$
GeV or smaller. The fissure is situated over the bisector, and it is due to 
a zero in the neutralino-neutralino-$Z$ coupling, {\it i.e.} a dynamical
suppression.

The branching ratio for the radiative decay 
$B(\chi^0_2\rightarrow\chi^0_1 \gamma)$ is shown in Fig.~\ref{M1M2Scan}c. We
see that the second lightest neutralino one-loop generated decay 
$\chi^0_2\rightarrow\chi^0_1 \gamma$ can dominate in a wide zone near
the bisector $M_1\sim M_2$. The reason for this possibility is that both
decay rates decrease sharply in parallel fractures but slightly displaced.
One of the fractures due to a zero in the $\chi^0\chi^0Z$ coupling 
(dynamical), and the other due to an eigenvalue degeneracy (kinematical). 
This is confirmed in 
Fig.~\ref{M1M2Scan}d, where we show both decay rates, the radiative decay 
branching ratio, and the $\chi^0\chi^0Z$ coupling as a function of $M_2$,
with constant values for $M_1=275$ GeV, $\mu=400$ GeV, $\tan\beta=10$,
and $\widetilde m=10^{4}$ GeV. We see that
the zero for $\Gamma(\chi^0_2\rightarrow\chi^0_1 f\overline f)$ coincides
with the zero for $\chi^0\chi^0Z$ coupling at $M_1 = M_2$, and that the
minimum for $\Gamma(\chi^0_2\rightarrow\chi^0_1 \gamma)$ coincides with
the point where $m_{\chi^0_2}-m_{\chi^0_1}$ is minimum, at $M_2\gsim M_1$.

In order to better understand the above result, it is instructive to see 
the neutralino mass matrix in the basis
$[-i\widetilde{\gamma},-i\widetilde{Z},\widetilde{H}_1,\widetilde{H}_2]$, 
where
\begin{equation}
\left[\begin{matrix} \widetilde\gamma \cr \widetilde Z \end{matrix}\right]=
\left[\begin{matrix} c_W & s_W \cr -s_W & c_W \end{matrix}\right]
\left[\begin{matrix} \widetilde B \cr \widetilde W \end{matrix}\right]
\,,\qquad\qquad
\left[\begin{matrix} \widetilde H_1 \cr \widetilde H_2 \end{matrix}\right]=
\left[\begin{matrix} c_\beta & -s_\beta \cr s_\beta & c_\beta 
\end{matrix}\right]
\left[\begin{matrix} \widetilde H_d \cr \widetilde H_u \end{matrix}\right]
\,.
\end{equation}
In this basis the mass matrix in eq.~(\ref{X0massmat}) looks as follows,
\begin{eqnarray}
{\bf M}_{\chi^0}&=&\left[\begin{array}{cc}
{\bf M}_{\chi^0}^{gg} & {\bf M}_{\chi^0}^{gh} \\
{\bf M}_{\chi^0}^{hg} & {\bf M}_{\chi^0}^{hh} 
\end{array}\right] \hspace{4mm} 
\end{eqnarray}
with ${\bf M}_{\chi^0}^{gh}=({\bf M}_{\chi^0}^{hg})^T$. The different 
submatrices are equal to,
\begin{eqnarray}
{\bf M}_{\chi^0}^{gg}&=&\left[\begin{array}{cc}
M_1c^{2}_{W} + M_{2}s^{2}_{W} & (M_{2}-M_{1})s_{W}c_{W}  \\
(M_{2}-M_{1})s_{W}c_{W} & M_1s^{2}_{W} + M_{2}c^{2}_{W}  
\end{array}\right]
\nonumber \\
{\bf M}_{\chi^0}^{hh}&=&\left[\begin{array}{cc}
\mu s_{2\beta} & -\mu c_{2\beta}\\
 -\mu c_{2\beta} & -\mu s_{2\beta}
\end{array}\right]
\label{X0gghh}
\end{eqnarray}
for the blocks in the diagonal, and
\begin{equation}
{\bf M}_{\chi^0}^{gh}= \frac{v}{2}\left[\begin{array}{cc}
  c_{\beta}(\tilde{g}_{d}s_{W}- \tilde{g}'_d c_{W}) + 
 s_{\beta}(\tilde{g}_{u}s_{W}-\tilde{g}'_u c_{W}) & 
  s_{\beta}(\tilde{g}_{d}s_{W}- \tilde{g}'_{d} c_{W}) - 
 c_{\beta}(\tilde{g}_{u}s_{W}-\tilde{g}'_u c_{W})  \\
  c_{W}(c_{\beta}\tilde g_d + s_{\beta}\tilde g_u) +
  s_{W}(c_{\beta}\tilde g'_d + s_{\beta}\tilde g'_u) & 
  c_{W}(s_{\beta}\tilde g_d - c_{\beta}\tilde g_u) + 
  s_{W}(s_{\beta}\tilde g'_d- c_{\beta}\tilde g'_u)  
\end{array}\right]
\label{X0gh}
\end{equation}
for the off diagonal block. This neutralino mass matrix reduces to its 
analogous expression in the MSSM if we neglect the running from $\widetilde m$ 
and the weak scale:
\begin{equation}
{\bf M}_{\chi^0}^{gh} \longrightarrow
\frac{v}{2}\left[\begin{matrix}
0 & 0 \cr g/c_W & 0
\end{matrix}\right]
\qquad {\mathrm{as}}\qquad
\widetilde m\longrightarrow m_{weak}\,.
\label{HGmixing}
\end{equation}
In the MSSM case, the direct mixing between photino and 
higgsinos vanishes, but 
a direct coupling between zino and one of the higgsinos remains. This implies 
that in general the lightest neutralino has a non vanishing component of higgsino, which 
in turn translates into a non vanishing $\chi^0\chi^0Z$ coupling. In this way, 
the photino will decouple from higgsinos in the region $M_1 \sim M_2$, as seen 
from eq.~(\ref{X0gghh}), and the decay rate 
$\Gamma(\chi^0_2\rightarrow\chi^0_1 f\overline f)$ vanishes also, making the
decay $\chi^0_2\rightarrow\chi^0_1 \gamma$ the dominant one. In SS the mechanism 
is similar, but modified by RGE effects.

As we mentioned, in the dynamical suppression region where $M_1\sim M_2$ the 
decay mode $\Gamma(\chi^0_2\rightarrow\chi^0_1 f\overline f)$ is suppressed 
because the $Z$ coupling to the photino is absent, thus, when the LSP is 
nearly photino, the $Z\chi^0_1\chi^0_2$ coupling is nearly zero. 
In the MSSM this region 
does not exactly coincides with the kinematical suppression region where 
$m_{\chi^0_1}\sim m_{\chi^0_2}$ due to the remanent higgsino-gaugino mixing 
seen in eq.~(\ref{HGmixing}). In this case, phase space is small, and 
$\chi^0_2$ may be forced to decay into light mesons. 
\begin{figure}[!hhh]
\begin{center}
\includegraphics[angle=90,height=7.0cm]{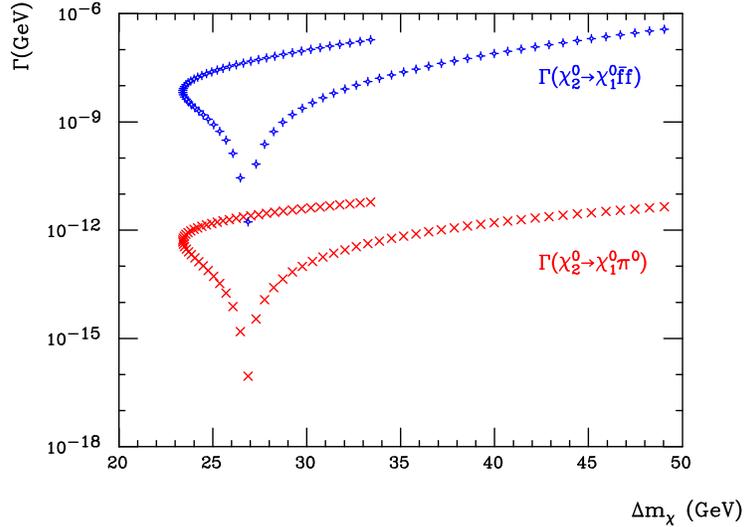}
\vspace{-3ex}\caption{Decay rate for $\chi^0_2\longrightarrow\chi^0_1\pi^0$
in comparison to the decay rate for 
$\chi^0_2\longrightarrow\chi^0_1 f\overline f$, as a function of the mass
difference $\Delta m_\chi=m_{\chi_2}-m_{\chi_1}$.
}
\label{pion}
\end{center}
\end{figure}
The situation is similar in Split Supersymmetry where the difference lies in 
the fact that in SS RGE effects separate further the regions where 
$Z\chi^0_1\chi^0_2$ coupling and $\Delta m_\chi$ vanish, as indicated by
the higgsino-gaugino mixing in eq.~(\ref{X0gh}). In Fig.~\ref{pion} we have 
plotted the decay rate $\Gamma(\chi^0_2\longrightarrow\chi^0_1 f\overline f)$ 
and the (included in the former) decay rate 
$\Gamma(\chi^0_2\longrightarrow\chi^0_1\pi^0)$ as a function of 
$\Delta m_\chi$. The independent variable we are varying is $M_2$, exactly as 
in Fig.~\ref{M1M2Scan}d. Both decay rates vanish at the point where the 
$Z\chi^0_1\chi^0_2$ coupling is null, but at that point $\Delta m_\chi$
is not zero. Indeed, in Fig.~(\ref{pion}) we have $\Delta m_\chi=26.8$ GeV 
for $\widetilde m=10$ TeV, while RGE effects changes it to 
$\Delta m_\chi=29.2$ GeV for $\widetilde m=M_{GUT}$. Therefore, the photon in 
the decay $\chi^0_2\rightarrow\chi^0_1 \gamma$ will have enough energy to be 
easily detected. Note that the two branches in each decay are defined by the
sign of $M_2-M_1$.

As we discussed, in Split Supersymmetry the mechanism is analogous to the MSSM, 
but the details are modified by the Renormalization Group Equations effects. 
Indeed, the remaining higgsino component of the lightest neutralino in the 
case $M_1 = M_2$ is controlled by the SS scale $\widetilde m$ via the RGE 
effects on the different couplings.
\begin{figure}[!hhh]
\begin{center}
\includegraphics[angle=0,height=7.0cm]{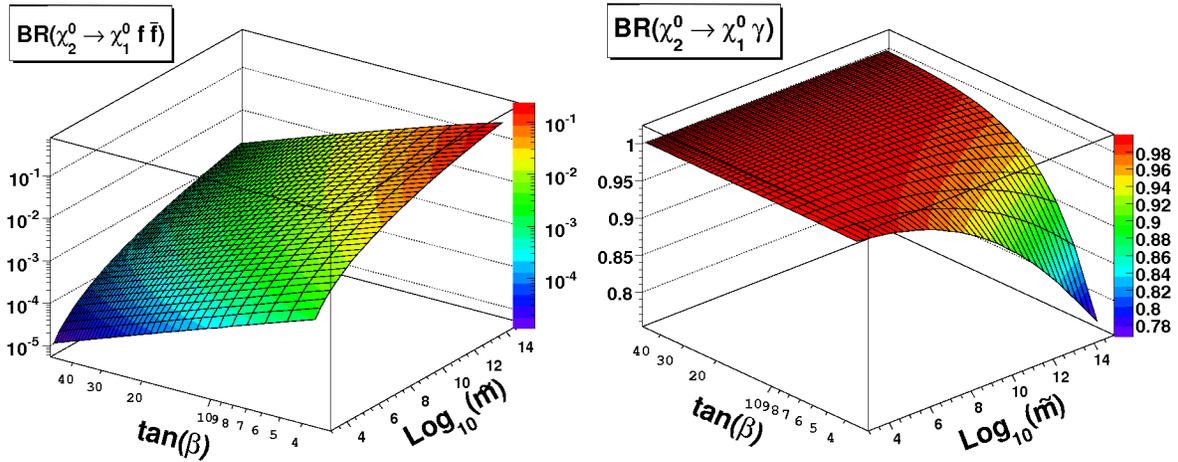}
\vspace{-3ex}\caption{$\chi^0_2$ decay modes as a function of the Split Susy 
scale $\widetilde{m}$ and $tan\beta$ for the scenario where $M_{1} = M_{2}$.}
\label{BR_1L_TL_mss_tbeta}
\end{center}
\end{figure}
This can be seen in Fig.~\ref{BR_1L_TL_mss_tbeta} where we have the $\chi^0_2$ 
branching ratios dependence on $\tan\beta$ and the SS scale $\widetilde m$,
with $\mu=400$ GeV and $M_1=M_2=275$ GeV. The dependence
on $\tan\beta$ is relatively mild, as it is the dependence on $\widetilde m$. 
But in comparison to other observables, the dependence on the SS scale is very
important, because in this scenario a measurement of the $\chi^0_2$ branching
ratios could yield valuable information on the SS scale, otherwise difficult to 
extract from experiments. In the right frame we see that 
$B(\chi^0_2\rightarrow\chi^0_1 \gamma)$ dominates over a large region, with a 
small decrease at small $\tan\beta$ and large $\widetilde m$. In the left frame 
we have $B(\chi^0_2\rightarrow\chi^0_1 f\overline f)$, with values that go from
$10^{-5}$ up to $0.2$. Clearly, a measurement of the branching ratio can 
give valuable information on the split supersymmetric scale.

\newpage
\section{Conclusion} 

We have calculated the decay rates and branching ratios of the second
lightest neutralino in Split Supersymmetry, and concentrate in the radiative 
decay $\chi^0_2\rightarrow\chi^0_1 \gamma$. We compared our results with 
mSUGRA, finding agreement when the scalar mass parameter is very large, 
$m_0\sim1$ TeV, where $B(\chi^0_2\rightarrow\chi^0_1 \gamma)\sim10^{-3}$. 
Small differences remain due to RGE effects, which increase with larger 
split supersymmetric scale $\widetilde m$. For larger values of 
$m_0$ comparison is not possible since large squark masses 
in quantum corrections destabilize the EWSB in direct mSUGRA calculations.

In general models, the possibility that $m_{\chi^0_2}$ is not much different 
than the mass of 
the LSP is experimentally challenging. This is because the decay products 
that can be detected, photons in the case of 
$\chi^0_2\rightarrow\chi^0_1 \gamma$ and fermions in the case of 
$\chi^0_2\rightarrow\chi^0_1 f\overline f$, are soft and specialized analysis 
have to be done with the data. 
Nevertheless, in our model the radiative decay dominates in a region where 
$\Delta m_\chi$ is large enough to produce an energetic photon.
We focus on the region $M_1\sim M_2$, where
this possibility is realized and show that the decay 
$\chi^0_2\rightarrow\chi^0_1 \gamma$ is dominant in a wide band around the
bisector $M_1=M_2$. Furthermore, in this region a measurement of the branching 
ratios $B(\chi^0_2\rightarrow\chi^0_1 \gamma)$ and
$B(\chi^0_2\rightarrow\chi^0_1 f\overline f)$ can give information on the value
of the split supersymmetric scale $\widetilde m$.

\begin{acknowledgments}
{\small 
We are indebted to Dr.~Pavel Fileviez-P\'erez for his insight in the early
stages of this work. We are thankful to Dr.~Benjamin Koch for useful 
comments. This work was partly founded by Conicyt and Banco Mundial grant 
``Anillo Centro de Estudios Subat\'omicos'', by Conicyt's ``Programa de 
Becas de Doctorado'', and by VRAID-PUC fellowships.
}  
\end{acknowledgments}

\newpage

\end{document}